\documentstyle[12pt,epsf]{article}

\advance\hoffset by -1.1truecm
\advance\voffset by -1.0truecm

\def\be{\begin{equation}}
\def\ee{\end{equation}}
\def\ba{\begin{eqnarray}}
\def\ea{\end{eqnarray}}
\def\v#1{\vert #1 \rangle}

\newcommand{\ord}{{\cal O}^2(\beta,\beta^\prime)}
\newcommand{\x}{{\bf x}}
\newcommand{\p}{{\bf p}}
\newcommand{\A}{{\cal A}}

\newcommand{\sn}{\smallskip\newline}
\newcommand{\mn}{\medskip\newline}

\newcommand{\mbo}{{\mbox{ }}}

\def\sec#1{\vskip0.5cm { \large \bf {\flushleft #1} \rm }\mn }

\begin{document}

\title{Nonpointlike Particles in Harmonic Oscillators}

\author{Achim Kempf\thanks{ Research Fellow of Corpus 
Christi College in the University of Cambridge}\\ 
Department of Applied Mathematics \& Theoretical Physics\\
University of Cambridge, Cambridge CB3 9EW, U.K.\\
 {\small Email: a.kempf@amtp.cam.ac.uk}}

\date{}
\maketitle

\vskip-8cm
{\tt $\mbo$ \hskip11.1cm DAMTP/96-39}\newline
{\tt $\mbo$ \hskip11.72cm  hep-th/9604045}\rm
\vskip7.7cm

\begin{abstract}
Quantum mechanics ordinarily describes particles as being pointlike,
in the sense that the uncertainty $\Delta x$ can, in principle, 
be made arbitrarily small. It has been shown that suitable correction
terms to the canonical commutation relations induce a finite
lower bound to spatial localisation. Here, we perturbatively 
calculate the corrections to the energy levels of an in this 
sense nonpointlike particle in isotropic harmonic oscillators.
Apart from a special case the degeneracy of the energy levels 
is removed. 
\end{abstract}
\sec{1. Introduction}
It has been shown that certain small corrections to the canonical
commutation relations yield an interesting new short distance 
structure, characterised by a finite minimal uncertainty $\Delta x_0$. 
The algebraic and functional analytic structure underlying minimal 
uncertainties in positions (and/or in momenta) first appeared in
\cite{ixtapa,ak-jmp-ucr}, and recent studies are e.g. 
\cite{ft}-\cite{ak-hh-1}.
The approach originated \cite{ak-lmp-bf,ak-jmp-bf}
in the field of quantum groups \cite{sm-book}, 
which is related to noncommutative geometry \cite{connes}.
Part of the motivation to investigate 
a possible minimal uncertainty $\Delta x_0$ is related to
string theory and quantum gravity where this type of 
short distance behaviour has been suggested to arise 
at the Planck scale, see e.g. \cite{townsend}-\cite{maggiore}. 
For a review see \cite{garay}. On the other hand, examples of 
quanta which cannot be localised to a point are also 
e.g. nucleons, quasiparticles  
or various collective excitations. Our aim here is to investigate
whether the new ansatz, with suitably adjusted scales, 
may also serve
for an effective low energy description of such nonpointlike particles.
To this end we consider a nonpointlike particle 
in a $d$-dimensional isotropic harmonic oscillator.
The special case of the 
1-dimensional oscillator was solved in \cite{ak-gm-rm-prd}.
Here, the application of different methods allows us to 
perturbatively calculate the spectra for the general case
where we find, as a new effect, a characteristic
splitting of the ordinarily degenerate energy levels.
\sec{2. Heisenberg algebra}
The associative Heisenberg algebra generated by $\x$ and $\p$
with commutation relations ($\beta >0$ assumed small) 
$[\x,\p]=i\hbar(1+\beta \p^2)$ yields the uncertainty relation
$\Delta x \Delta p \ge \hbar/2 
(1+\beta (\Delta p)^2 + \beta \langle \p^2\rangle)$
which is readily checked
to imply a minimal uncertainty in positions 
$\Delta x_0 = \hbar \sqrt{\beta}$.
The functional analytic structure, and the properties
of the states of then maximal localisation are discussed in detail in
\cite{ixtapa,ak-jmp-ucr,ak-gm-rm-prd,ak-hh-1}.
\sn
In $d$ dimensions the ansatz
\be
[\x_i,\p_j] = i\hbar \Theta_{ij}(\p)
\ee
yields a minimal uncertainty $\Delta x_0>0$ for appropriate choices
of symmetric $\Theta$.
We will leave momentum space `classical', i.e.
$[\p_i,\p_j]=0$. The Jacobi identity and the requirements $\x_i =\x_i^*,
\p_i=\p_i^*$ then uniquely determine
the commutation relations for the now in the generic case
noncommutative position operators:
\be
[\x_i,\x_j] = i\hbar \{\x_a,\Theta^{-1}_{ar}\Theta_{s[i}\Theta_{j]r,s}\}
\label{xxf}
\ee
For simplicity we adopt a geometric notation with $f_{,s}$
standing for $\partial_{p_s}f$ and where
repeated indices are summed over.
The associative Heisenberg algebra $\A$ finds a Hilbert space 
representation e.g. on momentum space through:
\be
\p_i.\psi(p) = p_i \psi(p)
\ee
\be
\x_i.\psi(p) = i\hbar (1/2 \Theta_{ai,a} + \Theta_{ai}\partial_{p_a}) \psi(p)
\ee
For more details on the geometric structure 
see \cite{banach}.
Assuming a rotationally isotropic situation and 
the minimal uncertainty $\Delta x_0$ to be 
small, we here only consider the lowest order
correction terms to the canonical commutation relations 
\be
\Theta_{ij}(p) = \delta_{ij} + \beta \delta_{ij}\p^2 +\beta^\prime\p_i\p_j
\label{e5}
\ee
where $\p^2 := \sum_{i=1}^d\p_i\p_i$, and where
$\beta,\beta^\prime>0$ are assumed small of the first order.
The same mechanism as in the one-dimensional case
yields from the corresponding uncertainty relations
$\Delta x_i \Delta p_i
\ge \vert \langle [\x_i,\p_j]\rangle\vert/2$ an isotropic
(i.e. $\Delta x_0{}_i =\Delta x_0{}_j, \forall i,j$)
minimal uncertainty $\Delta x_0$ (dropping the index $i$):
\be
\Delta x_0= \hbar \sqrt{\beta d+\beta^\prime} 
\label{dx0}
\ee
The commutation relations in the  
Heisenberg algebra $\A$ then read ${[ \p_i, \p_j ]} = 0$, and:       
\begin{eqnarray}
 {[ \x_i, \p_j ]} & = & i\hbar (\delta_{ij} +\beta\delta_{ij} \p^2
+ \beta^\prime \p_i \p_j) + {\ord} 
\\
 {[ \x_i,\x_j ]} & = & 
 i\hbar(\beta^\prime/2-\beta)(\{\x_i,\p_j\}-\{\x_j,\p_i\}) +{\ord}
\end{eqnarray}
In the momentum representation, where 
\be
\x_i.\psi(p) = i\hbar\left[\left(\beta+\beta^\prime\frac{d+1}{2}
\right)p_i
+(\delta_{ia} +\beta \delta_{ia}p^2+\beta^\prime p_ip_a)\partial_{p_a}
\right]\psi(p)
\ee
we have to this order:
\be
[\x_i,\x_j].\psi(p) = \hbar^2 (\beta^\prime-2\beta)(p_i\partial_{p_j}-
p_j\partial_{p_i})\psi(p) +\ord
\label{xc}
\ee
In the case $\beta = \beta^\prime/2$ the $\x_i$
are commutative, which may therefore be considered a preferred choice
of parameters. The framework is then translation invariant in the
sense that $\p_i\rightarrow \p_i,\mbo \x_i\rightarrow \x_i + \alpha_i$
defines an algebra homomorphism of $\A$. More generally, this feature
holds for any $\Theta$ that obeys (from Eq.\ref{xxf}) 
$\Theta_{ia}\partial_{p_i}\Theta_{bc} = 
\Theta_{ib}\partial_{p_i}\Theta_{ac}$. 
In the symmetric case $\Theta_{ij}:=\delta_{ij}f(p^2)+g(p^2)p_ip_j$
this condition is $g=2 f f^\prime(f-2p^2f^\prime)^{-1}$, where we may
choose e.g. $f:=e^{\beta p^2}$ to recover to first order
Eq.\ref{e5} with $\beta^\prime=2\beta$.
\sn
Recall, that generally 
$Q\v{\lambda}=\lambda\v{\lambda}\Rightarrow (\Delta Q)_{\v{\lambda}}=0$. 
Indeed, due to $\Delta x_0>0$, for $\A$ there no longer exists
a spectral representation of the $\x_i$,
see \cite{ixtapa,ak-jmp-ucr}. Therefore, in \cite{ak-gm-rm-prd}
the concept of quasi-position
representation has been introduced. The quasi position
wave function of a state $\v{\psi}$ is $\psi(\xi) :=
\langle\phi^{ml}_{\xi}\vert\psi\rangle$ where the 
$\vert \phi^{ml}_\xi\rangle$ are states of now maximal spatial 
localisation around positions $\xi$ (replacing position
eigenstates). 
In \cite{ak-hh-1} 
the concept has been extended to include a possible $\Delta p_0>0$.
\sn
We stress that corrections to the commutation 
relations necessarily induce new physical features,
which could not alternatively be described by keeping the
ordinary commutation relations and adding corrections to
Hamiltonians. Although a conventional `point particle' 
system can have the same energy spectrum,
for a suitably chosen Hamiltonian, the full set of physical predictions 
will generally differ. 
This is because for all predictions,
such as energy spectra, transition amplitudes and expectation values
to match,
systems must be related by a unitary transformation. However,
unitary transformations are commutation relations preserving.
\sn
Nevertheless, 
in Hilbert space representations of the generalised commutation relations
the usual perturbative techniques for the
calculation of eigenvalues of hermitean operators, such as
Hamiltonians, are still applicable.
\sec{3. The Harmonic Oscillator}
A low energy approximation for most kinds of oscillations is
a $d$-dimensional harmonic oscillator, which we here for simplicity 
choose isotropic:
\be
H := \sum_{i=1}^d \left(\frac{\p_i^2}{2m} + 
\frac{m \omega^2 \x_i^2 }{2}\right)
\ee
The Hamiltonian acts on momentum space as
\be
H.\psi(p)  = \left[\frac{p^2}{2m} -\frac{m\omega^2\hbar^2}{2}
\sum_{i=1}^d\left( \left(\beta + \beta^\prime\frac{d+1}{2}\right) p_i+
\partial_{p_i}+\beta p^2\partial_{p_i} +\beta^\prime p_i 
p_a \partial_{p_a} \right)^2\right]\psi(p)
\ee
which is, to first order in $\beta,\beta^\prime$:
\begin{eqnarray}
H.\psi(p) = \bigg[\frac{p^2}{2m} -\frac{m\omega^2\hbar^2}{2}\bigg(
\beta d &+& \mbo\beta^\prime (d+d^2)/2  + 
(4\beta+(2+2d)\beta^\prime)p_i\partial_{p_i} \nonumber\\
 & & +\mbo  2\beta p^2\partial^2+2\beta^{\prime} 
 p_ip_a\partial_{p_i}\partial_{p_a}
+\partial^2\bigg)\bigg] \psi(p)
\label{Hp}
\end{eqnarray}
Since we are dealing with harmonic oscillators it is convenient to
further transform into a Fock representation 
where $\v{\psi} = \psi(a^\dagger)\v{0}$.
The multiplication and differentiation operators 
$p_i$ and $\partial_{p_j}$ can be represented through
\begin{eqnarray}
p_j.\v{\psi} &=& i (m\omega \hbar/2)^{1/2} (a^\dagger_j -a_j)\v{\psi}
\\
\partial_{p_j}.\v{\psi} 
&=& -i (2m\omega \hbar)^{-1/2} (a^\dagger_j +a_j)\v{\psi}
\end{eqnarray}
where $a_i a^\dagger_j-a^\dagger_j a_i = \delta_{ij}$, so that  
\begin{eqnarray}
p_i\partial_i &=& -\frac{d}{2} + 
\mbox{NN-terms} \label{e1}
\\
p^2 \partial^2 &=& 
- N^2 -N(d+1)-\frac{d(d+2)}{4} +
\frac{1}{2}\sum_{i,j=1}^d a^2_ia_j^\dagger{}^2 + \mbox{NN-terms}
\\
p_ip_j\partial_i\partial_j &=&
N+\frac{d(d+4)}{4} -\frac{1}{2}
\sum_{i,j=1}^d a^2_ia_j^\dagger{}^2 + \mbox{NN-terms}
\label{e3}
\end{eqnarray}
where $N_i:=a^\dagger_ia_i$, $N:=\sum_{i=1}^dN_i$, and where
NN-terms are terms that contain nonequal numbers of
raising and lowering operators.
Substituting Eqs.\ref{e1}-\ref{e3} in Eq.\ref{Hp} yields for the
action of $H$ on Fock space:
\begin{eqnarray}
H \v{\psi} & = & \bigg[ \hbar \omega\left(N+\frac{d}{2}\right) 
\nonumber \\
  &  & \quad + \mbo m\omega^2\hbar^2\left(
\beta N^2 + \left(\beta (d+1)-\beta^\prime\right) N +
(\beta d(d+4)-3 \beta^\prime d )/4\right)\nonumber \\
  &   & \quad -\mbo m\omega^2\hbar^2
  \frac{\beta-\beta^\prime}{2} 
\sum_{i,j=1}^d a^2_ia_j^\dagger{}^2 + \mbox{NN-terms}
\bigg]\mbo \v{\psi}
\label{Hg}
\end{eqnarray}
The natural length scale of the harmonic oscillator
is the inverse length in the exponent
of the Hermite functions: $(\hbar/m\omega)^{1/2}$. Let us 
replace the parameters $\beta$ and $\beta^\prime$ by more intuitive
dimensionless parameters $k,k^\prime$ which
measure the minimal uncertainty 
length scales associated with $\beta$ and $\beta^\prime$
in units of the length scale of the oscillator (see Eq.\ref{dx0}):
$k := \hbar \sqrt{\beta}/(\hbar/m\omega)^{1/2}, \mbo
 k^{\prime} :=  \hbar 
\sqrt{\beta^\prime}/(\hbar/m\omega)^{1/2}$,
i.e. we have $\beta= k^2/m\omega\hbar$, $\beta^\prime=
k^\prime{}^2/m\omega\hbar$ and thus, from Eq.\ref{dx0},
\be
\Delta x_0 = \sqrt{k^2 d+k^\prime{}^2} 
\mbo \sqrt{\frac{\hbar}{m\omega}}
\ee
so that
\begin{eqnarray}
H \v{\psi} & = & \hbar\omega \bigg[ N+\frac{d}{2}
+ k^2N^2 + \left(k^2 (d+1)-k^\prime{}^2\right)N+
\frac{k^2 d(d+4)-3 k^\prime{}^2 d}{4} \nonumber \\
  &   & \quad -\mbo 
  \frac{k^2-k^\prime{}^2}{2} 
\sum_{i,j=1}^d a^2_ia_j^\dagger{}^2 + \mbox{NN-terms}
\bigg]\mbo \v{\psi}
\label{Hk}
\end{eqnarray}
\sec{4. First order corrections to the spectra}
We read off from Eq.\ref{Hk} that
$H$ consists of a diagonal part with degenerate eigenvalues
(in more than one dimension), and a nondiagonal term
$\sum_{i,j=1}^da^2_ia_j^\dagger{}^2$ 
proportional to $(k^2-k^\prime{}^2)$. 
As a new effect, this nondiagonal term can lead to a
splitting of the normally $g(n,d)$-fold degenerate eigenvalues
$E_n$ of the $d$-dimensional isotropic harmonic oscillator.
We recall the degeneracy function: $
g(n,d) = \frac{(n+d-1)!}{n!(d-1)!}$. 
As is well known, in the calculation of 
the eigenvalues of $A := B+C$, for $A,B,C$ hermitean, the first order
perturbative corrections to degenerate eigenvalues of $B$ are
the eigenvalues of the perturbing matrix $C$ when restricted to
the corresponding eigenspaces. 
\sn
Thus, here the $g(n,d)$-fold degenerate energy levels $E_n$ split
into levels $E^\prime_{n_r}$:
\begin{eqnarray}
E_{n_r}^\prime(k,k^\prime) &=&
\hbar \omega\bigg( n+\frac{d}{2} 
+ \mbo 
k^2n^2 + k^2 (d+1)n-k^\prime{}^2n
+ \frac{k^2 d(d+4)-3 k^\prime{}^2 d }{4} \nonumber\\
 & & \qquad - \frac{k^2-k^\prime{}^2}{2}\mbo \times \mbo
\mbox{r'thEigenvalue}\left[\left(\sum_{i,j=1}^d 
a_i^2 a_j^\dagger{}^2\right)\bigg\vert_{ {\cal{H}}_n}\right]\bigg)
\label{Es}
\end{eqnarray}
where
$r=1,2,...,g(n,d)$
and where the eigenspaces ${\cal H}_n$ of 
the diagonal part of the Hamiltonian
are 
\be
{\cal H}_n := \mbox{span}\left\{ (r_1!\cdot ... \cdot r_d!)^{-1/2}
a_1^\dagger{}^{r_1}\cdot .... \cdot 
a_d^\dagger{}^{r_d}\v{0} \bigg\vert \sum_{i=1}^d r_i = n\right\}
\label{Hn}
\ee
The matrix elements of NN-terms vanish in ${\cal H}_n$, i.e.
for $m = \sum_{i=1}^d r_i = \sum_{i=1}^d s_i$ there holds
$\langle 0\vert a_1^{s_1}\cdot...\cdot a_d^{s_d} (\mbox{ NN-terms })
 a_1^\dagger{}^{r_1}...a_d^\dagger{}^{r_d}\v{0}
= 0$,
so that the NN-terms of Eq.\ref{Hk} do not contribute in Eq.\ref{Es}.
\sn
For the calculation of the eigenvalues of 
$\sum_{i,j=1}^da_i^2a_j^\dagger{}^2$ in ${\cal H}_n$
we can choose the ON-basis given in Eq.\ref{Hn}
to obtain the matrix elements ($n=\sum_{i=1}^d r_i =\sum_{i=1}^d s_i$):
\begin{eqnarray}
\langle r_1,...,r_d\vert \sum_{i,j}^d
a_i^2a_j^\dagger{}^2\vert s_1,...,s_d\rangle & = &
\sum_{i,j=1}^d \sqrt{(r_i+1)(r_i+2)(s_j+1)(s_j+2)}\nonumber\\
 & & \quad \quad \times \mbo \delta_{r_1,s_1}...
\delta_{r_i+2,s_i}...\delta_{r_j,s_j+2}...\delta_{r_d,s_d}
\end{eqnarray}
We begin with the one-dimensional case. For this case the momentum space 
Schr{\"o}\-ding\-er equation proved to be exactly solvable in terms of
hypergeometric functions, yielding to first order in $\beta$,
from Eqs.53,56,69 in \cite{ak-gm-rm-prd}:
\be
E^\prime_n= \hbar\omega(n+1/2) + m\omega^2\hbar^2\beta(n^2/2 + n/2 +1/4)
\ee
Indeed, we recover this result from Eqs.\ref{Hg},\ref{Es} 
as the special case 
$d=1$ (note
that $a^2a^\dagger{}^2=N^2+3N+2$ and that
$\beta +\beta^\prime$, corresponds to $\beta$
in \cite{ak-gm-rm-prd}).
\newline
For $d=2$, straightforward calculation now yields the
$g(n,2)=n+1$ eigenvalues
of $\sum_{i,j=1}^2a^2_ia_j^\dagger{}^2$ in ${\cal H}_n$.
If $n$ is odd, these can be put into the form 
$4 s (n +2 - s)$ for $s=1,...,(n+1)/2$ with all eigenvalues
two-fold degenerate. For $n$ even, $s$ runs $s=1,...,(n+2)/2$ with
the last eigenvalue nondegenerate.
Using Eq.\ref{Es} we therefore obtain the energy 
levels for $d=2$ (with the degeneracies
given within brackets):
\begin{eqnarray}
E^\prime_{n_r}(k,k^\prime) &=& 
\hbar\omega \bigg[n+1+k^2(n^2+3n+3)-k^\prime{}^2
\left(n+\frac{3}{2}\right)\\
 & & \quad \qquad
-(k^2-k^\prime{}^2)
\left\{\matrix{2\cdot 1 (n+2-1) & (2\times) \cr
2\cdot 2 (n+2-2) & (2\times) \cr
2\cdot 3 (n+2-3) & (2\times) \cr
\vdots &   \cr
(n+1)(n+3)/2 & (2\times) \mbox{ for $n$ odd} \cr
(n+2)^2/2 &\mbo (1\times) \mbox{ for $n$ even}
}\right.\nonumber
\end{eqnarray}
Of particular interest is the 3-dimensional oscillator.
To illustrate the calculation, consider e.g. 
the splitting of the second excited energy level $E_2$. We may choose
as an ON-basis of ${\cal H}_2$ (see Eq.\ref{Hn}): 
$e_1:=\v{2,0,0}, e_2:=\v{0,2,0}, e_3:=\v{0,0,2},
e_4:=\v{0,1,1}, e_5:=\v{1,0,1}, e_6:= \v{1,1,0}$, in which:
\be
\left(\sum_{i,j=1}^d a^2_i a_j^\dagger{}^2\right)\bigg\vert_{{\cal H}_2}
= \left(\matrix{16 & 2 & 2 & 0 & 0 & 0 
\cr 2 & 16 & 2 & 0 & 0 & 0 \cr 2 & 2 & 16 & 0 & 0 & 0 \cr
0 & 0 & 0 & 14 & 0 & 0 \cr
0 & 0 & 0 & 0 & 14 & 0 \cr  
0 & 0 & 0 & 0 & 0 & 14 } \right)
\ee
The eigenvalues are: $14,14,14,14,14,20$. 
Thus, the ordinarily 6-fold degenerate second excited level $E_2$ 
splits into two energy levels, one of which is
$5$-fold degenerate and one nondegenerate (see Eq.\ref{2nd}).
The calculation of the first few excited states shows the
systematics in the splitting of the levels:
\begin{eqnarray}
E^\prime_{0} &=&
\hbar\omega \left[ \frac{3}{2} + 
\frac{21 k^2-9k^\prime{}^2}{4} -(k^2-k^\prime{}^2)\cdot 3 \mbo
 (1\times)
\right]
\\
E^\prime_{1_r} &=&
\hbar\omega \left[ \frac{5}{2} + 
\frac{41 k^2-13 k^\prime{}^2}{4} 
-(k^2-k^\prime{}^2)\cdot 5 \mbo (3\times) \right]
\label{1st}
\\
E^\prime_{2_r} &=&
\hbar\omega \left[ \frac{7}{2} + \frac{69 k^2 -17k^\prime{}^2}{4}
-(k^2-k^\prime{}^2)\cdot\left\{
\matrix{7 \mbo (5 \times) \cr 10 \mbo (1 \times) }\right.\right]
\label{2nd}
\\
E^\prime_{3_r} &=&
\hbar\omega \left[ \frac{9}{2} + \frac{105 k^2 -21k^\prime{}^2}{4}
-(k^2-k^\prime{}^2)\cdot\left\{
\matrix{9 \mbo (7 \times) \cr 14 \mbo (3 \times) }\right.\right]
\label{3rd}
\\
E^\prime_{4_r} &=&
\hbar\omega \left[ \frac{11}{2} + \frac{149 k^2 -25 k^\prime{}^2}{4}
-(k^2-k^\prime{}^2)\cdot\left\{
\matrix{11 \mbo (9 \times) \cr 18 \mbo (5 \times)
\cr 21 \mbo (1 \times) }\right.\right]
\label{4th}
\\
E^\prime_{5_r} &=&
\hbar\omega \left[ \frac{13}{2} + \frac{201 k^2 -29
 k^\prime{}^2}{4}
-(k^2-k^\prime{}^2)\cdot\left\{
\matrix{13 \mbo (11 \times) \cr 22 \mbo (7 \times)
\cr 27 \mbo (3 \times) }\right.\right]
\label{5th}
\end{eqnarray}
\sec{5. The translation invariant case}
The splitting of the energy levels for 
the case $k^\prime{}^2 = 2 k^2$ is interesting
for the framework
is then translation invariant and 
the $\x_i$ commute (Eq.\ref{xc}).
(Note that there would be
 no splitting for $k^\prime{}^2=k^2$).
\begin{eqnarray}
E^\prime_{0} &=& \frac{3}{2} \mbo
\hbar\omega   + \frac{3}{4}\mbo 
(\Delta x_0)^2 m\omega^2 \quad (1\times)
\label{el0}
\\
E^\prime_{1_r} &=& \frac{5}{2} \mbo
\hbar\omega +  \frac{3}{4}
\mbo(\Delta x_0)^2 m\omega^2
 \quad (3\times)
\\
E^\prime_{2_r} &=& \frac{7}{2} \mbo
\hbar\omega   + 
\mbo(\Delta x_0)^2 m\omega^2\cdot
\left\{\matrix{63/20 \mbo (5 \times) \cr 15/4 \mbo 
(1 \times) }\right.
\\
E^\prime_{3_r} &=& \frac{9}{2} \mbo
\hbar\omega  + 
\mbo(\Delta x_0)^2 m\omega^2\cdot
\left\{ \matrix{99/20 \mbo (7 \times) \cr 119/20 \mbo 
(3 \times) }\right.
\\
E^\prime_{4_r} &=& \frac{11}{2} \mbo 
\hbar\omega  + 
\mbo(\Delta x_0)^2 m\omega^2 \cdot
\left\{ \matrix{143/20 \mbo (9 \times) \cr 171/20 \mbo (5 \times)
\cr 183/20 \mbo (1 \times) }\right.
\label{el4}
\\
E^\prime_{5_r} &=& \frac{13}{2} \mbo 
\hbar\omega  + 
\mbo(\Delta x_0)^2 m\omega^2 \cdot
\left\{ \matrix{195/20 \mbo (11 \times) \cr 231/20 \mbo (7 \times)
\cr 251/20 \mbo (3 \times) }\right.
\label{el5}
\end{eqnarray} 
Eqs.\ref{el0}-\ref{el5} give the first few levels for this case
 and $d=3$, expressed in terms of the then only free parameter
 $\Delta x_0$.
For a graph of the spectrum see Fig1.\mn
\epsfxsize=2.5in \centerline{\epsfbox{oscpic.eps}}
\vskip0.5cm
\centerline{\small \it Fig1: 
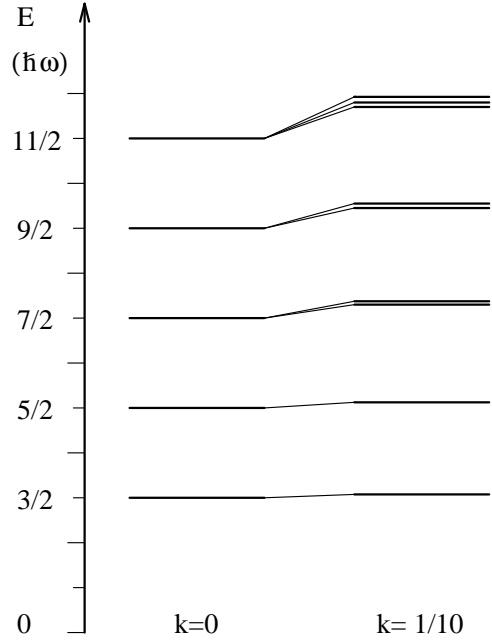
Splitting of the energy levels of the 3-dim isotropic 
harmonic }
\centerline{\small \it 
oscillator for a nonpointlike particle with
relative size $k=1/10$.}
\vskip0.4truecm
\sec{6. Discussion}
We are considering
oscillators for which a lowest order anharmonicity 
originates in the nonpointlike nature of the oscillating
particle. Instead of describing nonpointlike particles from
first principles,
such as nucleons in terms of their quark content, we here 
aim at an effective low energy 
description which accounts for nonpointlikeness
through the introduction of a finite $\Delta x_0$.
Our calculation of the lowest order corrections to the 
energy levels of an in this sense nonpointlike particle in an isotropic 
harmonic oscillator predicts the new effect of a 
characteristic fine splitting of 
the ordinarily degenerate energy levels.
\sn
While rotation invariance of the commutation relations 
allows two free parameters,
the additional requirement of translation
invariance (the case
$\beta^\prime = 2\beta, k^\prime{}^2=2k^2$), 
reduces these to one free parameter $\Delta x_0=
\hbar((d+2)\beta)^{1/2}
=k ((d+2)\hbar/m\omega)^{1/2}$.
With only a single parameter $\Delta x_0$ to fit,
it should in principle be possible to experimentally check
the applicability of the approach
for any given species of nonpointlike particles.
\sn
The new approach of course also allows to straightforwardly calculate 
arbitrary further details of the system, such as
wave functions, transition amplitudes or
probability amplitudes for spatially localised measurements,
in arbitrary dimensions and also for non-isotropic oscillators.
\sn
Concerning the possibility of a
 \it fundamental \rm $\Delta x_0>0$, which we 
mentioned in the beginning, it should be interesting to
apply similar perturbative methods to the case of the Hydrogen atom.
This should yield the relation between the
scale of an assumed non-pointlikeness of the electron and the scale
of the thereby caused effects on the Hydrogen spectrum. The 
high precision to which such effects can be experimentally excluded
could then yield an interesting upper bound for a possible 
fundamental non-pointlikeness
$\Delta x_0$ of the $e^-$. This issue is presently under investigation.
Recent studies on minimal uncertainties and regularisation in field theory
are \cite{ak-qft,ak-reg}. For related studies see e.g.
\cite{doplicher,grosse}.

\end{document}